\documentclass[prl,aps,twocolumn,superscriptaddress,showpacs]{revtex4}
\usepackage{graphicx}

\begin{document}
\title{The nature of the phase transition in dipolar fluids}

\author{J. M. Tavares}
\affiliation{Centro de F\'{\i}sica Te\'{o}rica e Computacional
da Universidade de Lisboa \\
Avenida Professor Gama Pinto 2, P-1649-003 Lisboa, Portugal}
\affiliation{Instituto Superior de Engenharia de Lisboa\\ Rua Conselheiro 
Em\'{\i}dio Navarro, 1, P-1949-014 Lisboa, Portugal}
\author{J. J. Weis}
\affiliation{Laboratoire de Physique Th\'eorique, Universit\'e 
de Paris XI, B\^atiment 210,
F-91405, Orsay Cedex, France}
\author{M. M. Telo da Gama}
\affiliation{Centro de F\'{\i}sica Te\'{o}rica e Computacional
da Universidade de Lisboa \\
Avenida Professor Gama Pinto 2, P-1649-003 Lisbon, Portugal}
\affiliation{Departamento de F\'{\i}sica, Faculdade de Ci\^encias da
Universidade de Lisboa, \\
Campo Grande, Lisboa, Portugal}

\date{today}

\begin{abstract}
Monte Carlo computer simulations of a quasi two dimensional (2D) dipolar
fluid
at low and intermediate densities indicate that the structure of the
fluid is well described by an ideal mixture of self-assembling clusters.
A detailed analysis of the topology of the clusters, of their
internal energy and of their size (or mass) distributions further suggests
that the system undergoes a phase transition from a dilute phase 
characterized
by a number of disconnected clusters to a condensed phase characterized
by a network or spanning (macroscopic) cluster that includes most of the
particles in the system.
\end{abstract}


\maketitle

\begin{figure} 
\begin{center}
\includegraphics[width=8cm]{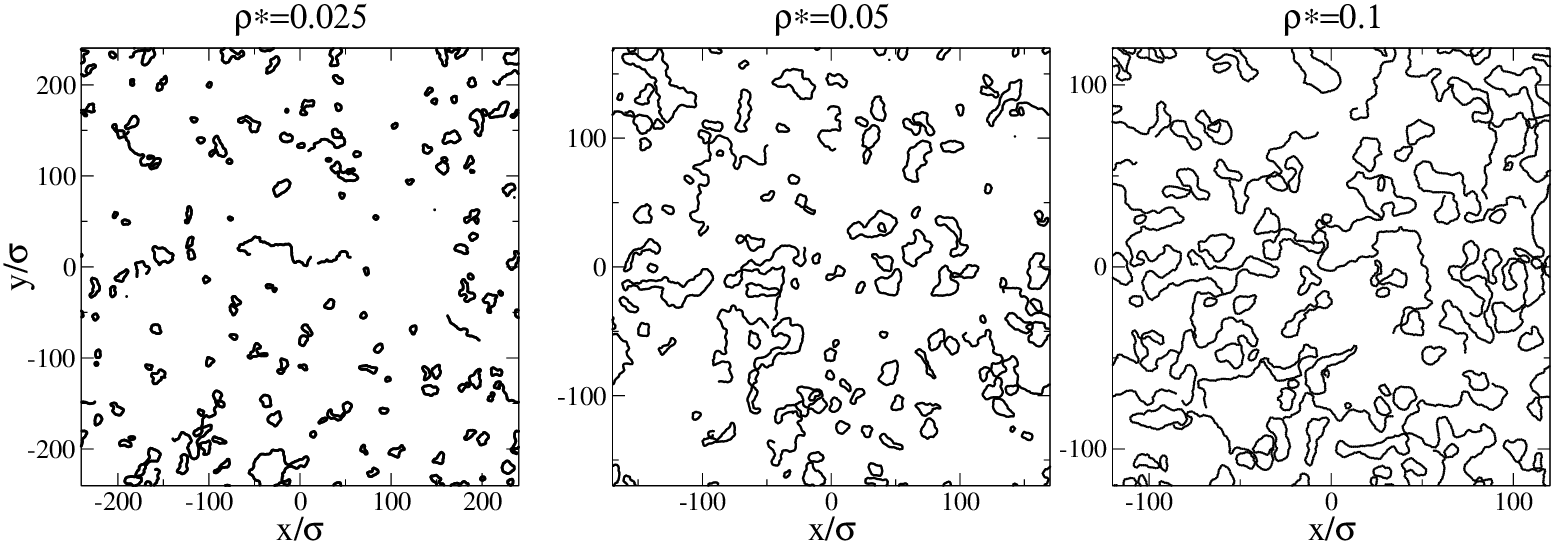} 
\end{center}
\caption{Typical equilibrium configurations of the quasi 2D 
DHS as obtained from MC simulations with $N_p=5776$ at $m^*=2.75$ and 
three reduced densities.} 
\label{fig1} 
\end{figure}

The condensation of simple fluids results from the free energy
balance of the high entropy gas and the low energy liquid phases.
This transition appears to be generic in simple fluids interacting
through isotropic intermolecular potentials that are repulsive at short
distances and attractive otherwise.
The dipolar hard sphere fluid (DHS) is a model where hard (or soft) 
spheres with an embedded central dipole interact through the 
dipole-dipole potential. As the average dipolar interaction between 
two dipoles (weighted by the Boltzmann factor) is attractive 
one may expect a phase behavior analogous to that of simple fluids. 
A recent calculation of the free energy of the DHS at several 
temperatures, using Monte Carlo (MC) simulations \cite{camp2000}, suggests 
the presence of an isotropic fluid-fluid transition at low densities, 
lending some support to the analogy with simple fluids. 
However, the structure of DHS at those low densities is drastically 
different. Numerical simulations of DHS 
\cite{simold1} for 
dipolar interaction strengths of the order of the thermal energy, 
have shown that the anisotropy of the 
dipolar potential promotes the formation of self-assembled aggregates 
(chains, rings and more complex clusters - see fig.~\ref{fig1}) 
in sharp contrast with 
the isotropic compact clusters observed in simple fluids. 
Moreover, the pair correlation function of DHS is strongly peaked at 
contact and the internal energy is nearly independent of the density at 
odds with the behavior of simple fluids. 
 
Association theories \cite{vanroij,sear,tavares99,tavaresrev,tavares2002}, 
that include the effect of cluster formation in the thermodynamics, 
describe well the slow variation of the internal energy with 
the density and the size (or mass) distribution of the clusters. The 
simplest theoretical approach (based on simulation results 
\cite{simold1,tavares99,tavares2002}) assumes that the only effect of the 
dipolar interactions is to drive 
cluster 
formation. Thus, the DHS is described as an ideal mixture of 
self-assembling clusters. 
 
The Helmholtz free energy density, $f$, of 
a system of volume $V$ (or area $A$ in 2D), 
at absolute temperature $T$ and with 
$N_p$ particles that self assemble to form non-interacting 
clusters 
of size $N$ $(1\le N\le N_p)$, is \cite{tavares2002,Sear2002} 
\begin{equation} 
\label{eq:freeensa} \beta f = \sum_{N=1}^{N_p}\phi(N) \left(\ln 
\phi(N) - 1 - \ln q(N) \right), 
\end{equation} 
where  $\beta\equiv(k_{\rm{B}}T)^{-1}$ ($k_{{\rm B}}$ is 
Boltzmann's constant).  $\phi(N)$ and $V q(N)$ are, respectively, 
the density and the partition function of clusters of size $N$. 
The free energy $f$ is minimized with respect to the densities 
$\phi(N)$, subject to a normalization condition $N_p/V = 
\sum_{N=1}^{N_p}N\phi(N)$, yielding the set of equations 
\cite{tavares2002}, 
\begin{equation} 
\label{eq:rhoNgen} \phi(N)= 
q(N) \exp\left(N\beta \mu(N_p,\rho) 
\right), 
\end{equation} 
where $\mu(N_p,\rho)$ is the chemical potential of a system with $N_p$ 
particles and density $\rho\equiv N_p/V$. In general 
\cite{desCloizeaux}, the partition function $q(N)$ may be written 
as 
\begin{equation} 
\label{eq:partf} q(N)= F(N) \exp(-\beta\lambda N), 
\end{equation} 
where $\lambda$ is the free energy per particle of an infinite 
cluster and $-\beta^{-1}\ln (V F(N))$ is a sublinear correction (in $N$) to 
that free energy. The substitution of eqs. (\ref{eq:rhoNgen}) and 
(\ref{eq:partf}) in the normalization condition results in,
\begin{equation} 
\label{eq:normal} 
\rho =\sum_{N=1}^{N_p} NF(N)\exp\left[\beta\left(\mu(N_p,\rho)
-\lambda\right)N\right], 
\end{equation} 
and defines implicitly $\mu(N_p,\rho)$. 
The phase behavior of the system is obtained from the equation of state 
$\mu(\rho)\equiv\lim_{N_p\to\infty} \mu(N_p,\rho)$, 
{\it {i.e.}} the thermodynamic limit of eq.~(\ref{eq:normal}). 
This limit depends crucially on the convergence the  
sum $\rho_t\equiv \sum_{N=1}^\infty N F(N)$. 
If $\rho_t$ diverges, then $\mu(\rho)$ is an analytic increasing 
function of $\rho$, bounded by $\lambda$ ($\mu(\rho) < \lambda$).
Consequently, when $\rho_t$ diverges, the system does exhibit any phase 
transition. 
On the other hand, if $\rho_t$ is finite, $\mu(\rho)$ converges 
non-uniformly to the function: 
$\mu(N_p\equiv\infty, \rho)$ (given through eq.~(\ref{eq:normal}))
if $\rho < \rho_t$; and $\lambda$ if $\rho \ge \rho_t$. 
Then, for finite $\rho_t$, 
$\mu(\rho)$ is singular at $\rho=\rho_t$, signaling a phase 
transition at this density: when $\rho<\rho_t$ the structure 
consists of small disconnected clusters ($\mu(\rho)<\lambda$) and when 
$\rho\ge\rho_t$ any excess particles (with respect to 
$\rho_t$) condense in an 'infinite', spanning 
cluster ($\mu(\rho)=\lambda$). 
This transition is similar to a variety of other 
transitions \cite{Sear2002} - lamellae formation 
in systems of disk-like micelles, emulsification failure in 
microemulsions, Bose-Einstein condensation, etc.- where 
condensation is not driven by the interactions between 'particles' (or 
aggregates). Previous applications of association theories to the 
DHS considered chain and ring formation only 
\cite{vanroij,sear,tavaresrev,tavares2002} and failed to predict a 
phase transition. Indeed, infinite chains and rings have the same 
configurational entropy \cite{desCloizeaux} and internal energy 
\cite{tavares2002} per particle. Their behavior is similar to that 
of self avoiding random walks with $F(N)\propto 
(N^{\gamma-1}+N^{-3+\alpha})$, where $\gamma$ and $\alpha$ are 
universal exponents known from polymer theory \cite{desCloizeaux}. 
Since $\gamma\ge1$, $\rho_t\to \infty$ and the absence of a phase 
transition follows from eq.~(\ref{eq:normal}). However, chains and 
rings are only the simplest self-assembled clusters in DHS 
\cite{tavares2002,tlusty2000} and the previous argument may not 
hold in general. In this letter we provide evidence for the 
occurrence of a phase transition in DHSs by analyzing the 
distribution functions of various types of clusters obtained from 
extensive Monte Carlo (MC) DHS simulations. Understanding the 
existence and nature of this phase transition is important 
for applications based on dispersions of ferromagnetic 
nanoparticles \cite{experiments}, where strong dipolar 
interactions are present, as well as for theoretical reasons. In 
fact, the interplay between cluster formation and condensation is 
a general problem, relevant in a variety of other theoretical contexts 
\cite{safran2002}. 
\begin{figure} 
\begin{center}
\includegraphics[width=7cm]{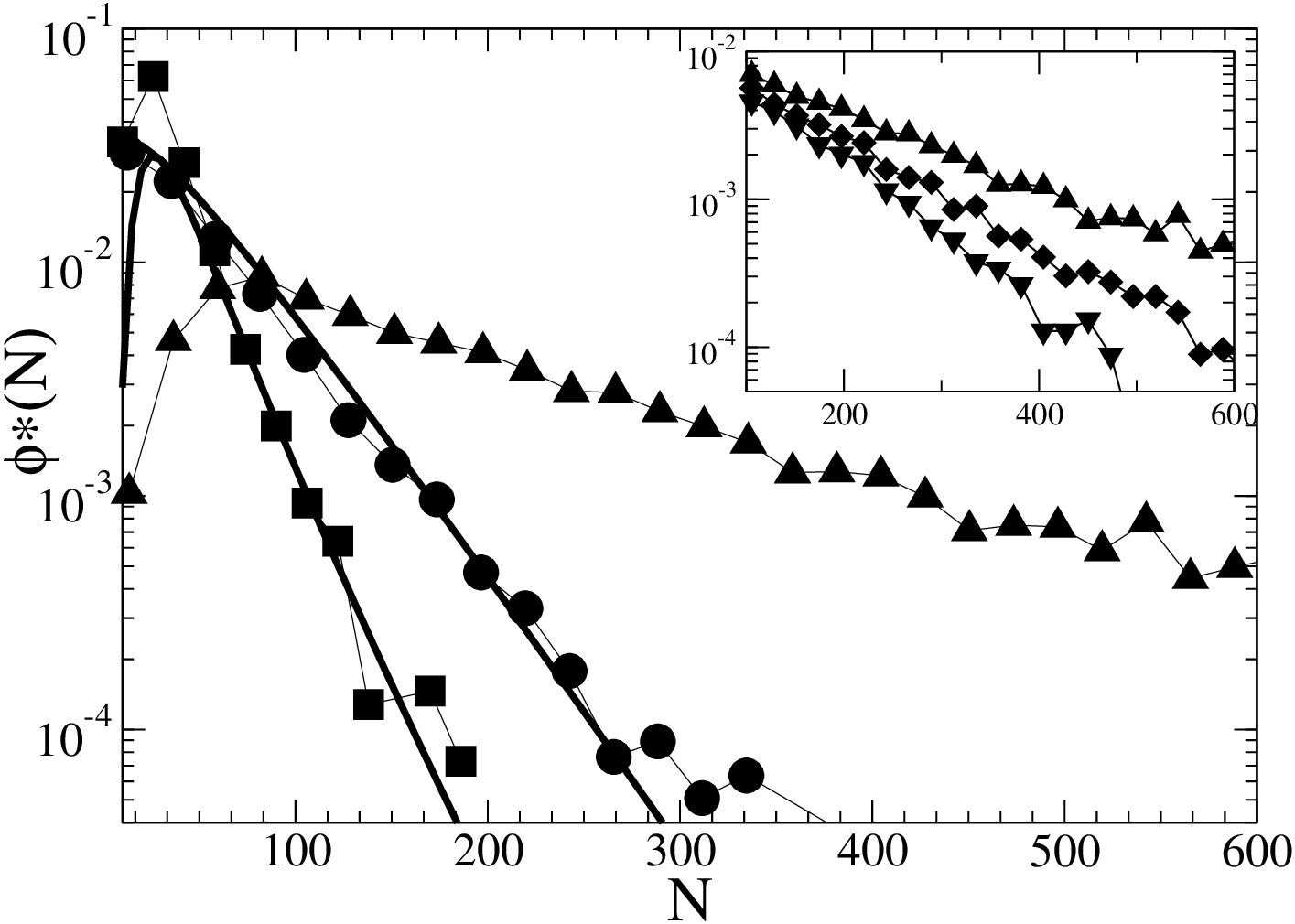} 
\vskip-0.4cm
\caption{ Mean reduced density of 
chains (circles, $\phi^*(N,2,0)$), rings (squares,
$\phi^*(N,0,0)$) and networks (triangles, $\phi^*_n(N)$) of size 
$N$, as obtained from simulations with $\rho^*=0.1$. The lines are 
the theoretical predictions for the chains and rings. Inset: mean 
reduced density of networks at $\rho^*=0.05$ (inverted triangles), 
$\rho^*=0.0625$ (diamonds) and $\rho^*=0.1$ (triangles). } 
\end{center}
\label{fig2} 
\end{figure} 
 
We have performed extensive MC simulations in the canonical ensemble for 
systems of hard spheres with diameter $\sigma$ and dipole strength $m$, 
interacting through the pair potential, 
\begin{equation} 
\label{eq:pot} U_{DHS}=U_{HS} -{m^2 \over r_{12}^3}\left [3({\hat 
\mu}_1 \cdot {\hat r}_{12}) ({\hat \mu}_2 \cdot {\hat r}_{12}) - 
{\hat \mu}_1 \cdot {\hat \mu}_2\right]. 
\end{equation} 
$r_{12}$ is the distance between spheres 1 and 2, $U_{HS}$ the 
hard-sphere potential ($=\infty$ if  $r_{12}<\sigma$, $0$ 
otherwise), ${\hat r}_{12}\equiv \frac{\vec r_2-\vec r_1}{r_{12}}$ 
the unit interparticle vector and ${\hat \mu}_1$, ${\hat \mu}_2$ 
the unit vectors in the direction of the dipole moments of spheres 
1 and 2, respectively. The centers of the spheres and their dipole 
moments were constrained to lie on the same plane, and thus the 
model is a quasi 2D dipolar hard sphere fluid \cite{tavares2002}. 
We used $N_p=5776$ particles and, for the isotherm $m^*\equiv 
{m}({ \sigma^3 k_{{\rm B}}T})^{-1/2} =2.75$, simulations were 
performed as described in \cite{tavares2002}, at reduced densities 
$\rho*\equiv\sigma^2\frac{N_p}{A}=0.025\,,0.03125\,,0.0375 
\,,0.05\,,0.075, 0.1, 0.15$ and $0.2$ ($A$ is the area of the 
simulation box). Between 100 (low densities) and 400 (high 
densities) equilibrium configurations were generated, resulting 
from 2 to 5 $\times 10^7$ MC cycles (i.e. attempts to rotate and 
move each of the $N_p$ dipolar spheres). 
 
Figure~\ref{fig1} shows snapshots of equilibrium configurations 
and evidence that the structure of this system is non trivial. The 
particles tend to aggregate in clusters of several sizes and 
topologies; some of the clusters exhibit linear aggregation only 
(chains and rings) and all the others (networks) have branches 
even if most of their particles are still linearly aggregated. 
This qualitative picture is quantified \cite{tavares2002} by 
defining a circle of diameter $r_c$ around each particle $i$: if 
there are one, two or more additional particles within this 
circle, then $i$ is an end, an interior or a junction particle, 
respectively \cite{safran2002}. Two particles belong to the same 
cluster if their separation is less than $r_c$ and the topology of 
each cluster is determined by the number of ends and junctions: 
rings have interior particles only, chains have two ends and no 
junctions, and networks have at least one junction. The cut off 
$r_c$ must be $\approx \sigma$ and in this work we took 
$r_c=1.15\sigma$. 
 
Both the total internal energy of the system and the internal energy of 
the clusters decrease 
$\approx 0.5\%$ when the density is increased from $\rho^*=0.025$ to 0.1. 
At all densities, the difference between these two energies 
is of the order of $0.5\%$, and well defined size 
distributions are observed even though the clusters 
break and recombine during a simulation run. Thus, as in 
previous works \cite{vanroij,sear,tavares99,tavaresrev,tavares2002}, 
the description of the system as an ideal mixture of 
self-assembling clusters is justified. The theory entailed in 
eqs.~(\ref{eq:rhoNgen},\ref{eq:partf},\ref{eq:normal}) is easily 
generalized to include $N$-clusters with different 
topologies. Each $N$-cluster is classified according to 
its number of ends, $n_1$, and junctions, $n_3$. 
The partition function of an $N$-cluster is now $V q(N,n_1,n_3)$, with, 
\begin{equation} 
\label{eq:pfgen} 
q(N,n_1,n_3) = F(N,n_1,n_3) \exp(-N\beta\lambda(x_1,x_3)), 
\end{equation} 
where $\lambda(x_1,x_3)$ is the free energy per particle of 
an infinite cluster with a fraction of ends 
$x_1\equiv n_1/N$ and a fraction of junctions 
$x_3\equiv n_3/N$, and $-\beta^{-1}\ln VF(N,n_1,n_3)$ is the 
sublinear correction to that free energy. 
The generalization of eq.~(\ref{eq:rhoNgen}) for the density 
of $N$-clusters with topology $(n_1,n_3)$ is, 
\begin{equation} 
\label{eq:rhoNgen2} 
\phi(N,n_1,n_3)= F(N,n_1,n_3) 
\exp(\beta N(\mu-\lambda(x_1,x_3))). 
\end{equation} 
The partition functions of chains and rings are approximated by 
those of self avoiding random walks  (SARW) with internal energies 
given by $E_{ch}(N) = -N\epsilon_0+2\epsilon_1$ (chains) and 
$E_r(N)= -N\epsilon_0+\epsilon_r/N$ (rings) \cite{tavares2002}. 
Here, $\epsilon_0$ is a ``bond'' energy, i.e. the internal energy 
per particle of an infinite chain or ring, $\epsilon_1$ is the 
energy cost of an end, and $\epsilon_r/N$ a long-range correction 
to the energy of a ring. The resulting size distributions are 
given, at fixed $m^*$, by $\phi (N,2,0) \propto 
N^{\gamma-1}\exp(\beta(\mu-\lambda_0)N)$ and $\phi(N,0,0) \propto 
N^{-3+\alpha} \exp(\beta(\mu-\lambda_0)N+\beta\epsilon_r/N)$, 
where $\lambda_0=\lambda(0,0)$. In fig.~\ref{fig2} we plot these 
predictions (with $\gamma=1.34$ and $\alpha=0.5$ of 2D 
SARW\cite{desCloizeaux}) for $\rho^*=0.1$ and the simulation 
results. The remarkable agreement found at this density is also 
observed at $\rho^*<0.1$. 

\begin{figure}
\begin{center} 
\includegraphics[width=7cm]{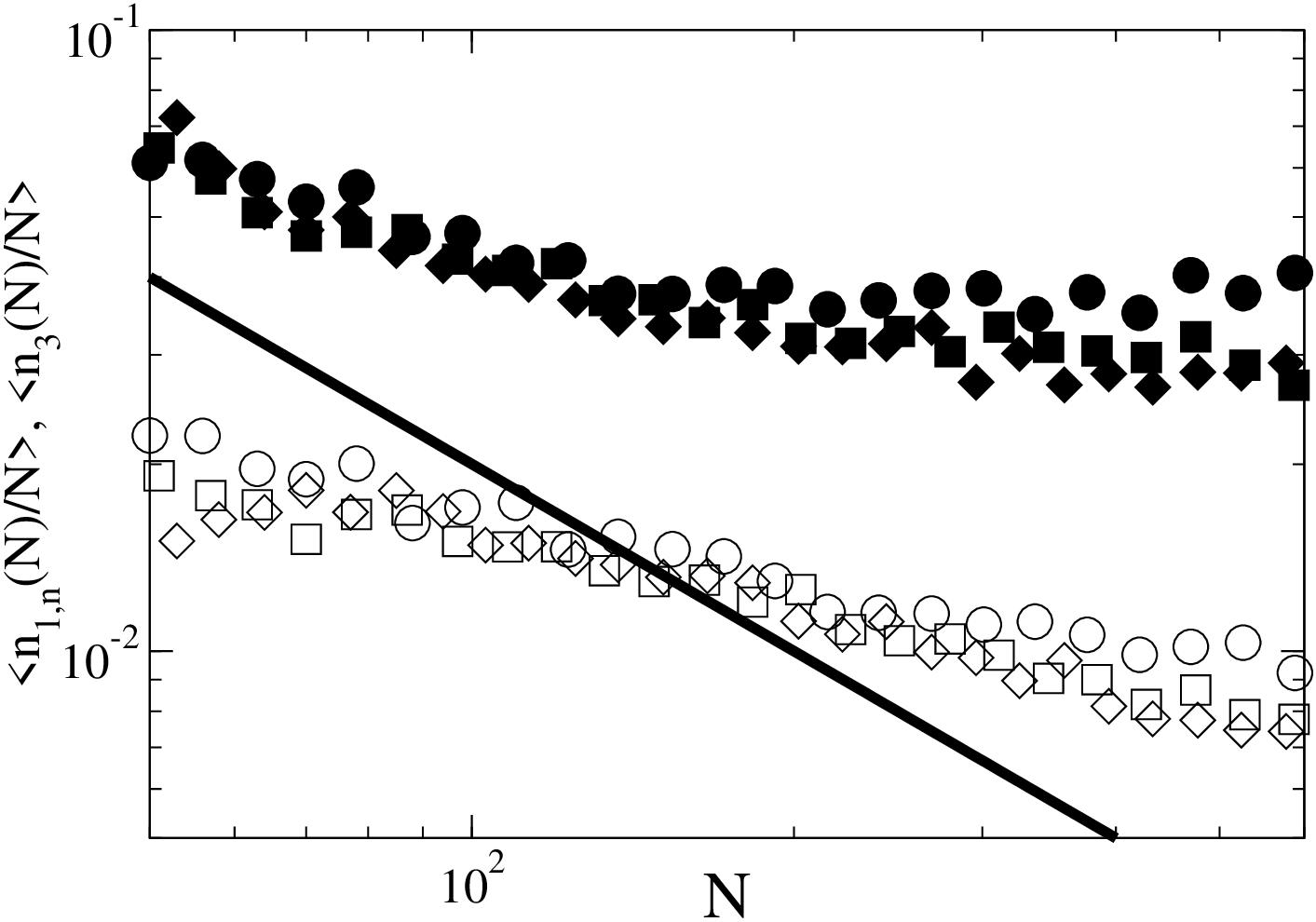} 
\vskip-0.5cm
\end{center}
\caption{Mean fraction of ends 
($<n_{1,n}>/N$, open symbols) and junctions ($<n_3>/N$, full 
symbols) in networks of size $N$, at $\rho^*=$ 0.1 (circles), 
0.075 (squares) and 0.0625 (diamonds). Simulation results for 
clusters with $N$ in the range $50-600$. Note that $<n_{1,n}>/N$ 
may be approximated by a power of $N$, with a negative exponent larger 
than $-1$ (cf. with the full line $2N^{-1}$, the fraction of ends in 
chains of size $N$). } 
\label{fig3} 
\end{figure}

The implementation of this approach 
requires the knowledge of $q(N,n_1,n_3)$ for networks (i.e. for 
$n_3\ne0$) and simulation results for the various cluster 
distribution functions. Unfortunately this is not possible at 
present, as the number of configurations of a network of $N$ 
monomers, with $n_1$ ends and $n_3$ junctions, is not known and 
simulations do not yield reliable distribution 
functions for clusters with arbitrary topologies. In order to 
proceed, we note that an increase in the number of junctions and 
ends of a $N$ cluster increases the internal energy, while it 
decreases the volume available to branches, decreasing the 
translational entropy of the network. However, the number of 
possible branched architectures increases, increasing the 
configurational entropy of the network. We may then conjecture 
that, at fixed $N$, the function $q(N,n_1,n_3)$ will exhibit a 
maximum for some $({\bar n}_1(N),{\bar n}_3(N))$. If, in the 
spirit of the saddle-point approximation, only this maximum is 
considered the mean density of networks of size $N$, $\phi_n(N)$, 
is, 
\begin{equation} 
\label{eq:rhonet} 
\phi_n(N)=F(N,{\bar n}_1,{\bar n}_3) 
\exp(\beta N(\mu-\lambda({\bar x}_1,{\bar x}_3))), 
\end{equation} 
where ${\bar x}_1\equiv{\bar n}_1(N)/N$ and ${\bar x}_3\equiv{\bar 
n}_3(N)/N$. Substituting this (and the corresponding expressions 
for chains and rings) in eq.~(\ref{eq:normal}) yields the 
equation of state. Further analysis requires the knowledge of the 
functions $F$, ${\bar n}_1(N)$ and ${\bar n}_3(N)$. In 
fig.~\ref{fig3} we plot the mean fraction of ends and junctions of 
$N$-networks, at three different densities, obtained from the 
simulations. Approximating ${\bar n}_1(N)$ and ${\bar n}_3(N)$ by 
these quantities, one obtains, in the large $N$ limit, ${\bar 
n}_3(N)\approx c_3 N$, where $c_3$ is constant (if a slight 
dependence on $\rho^*$ is neglected) and ${\bar n}_1 \approx c_1 
N^{\eta}$, with $c_1$ and $\eta$ constant. Comparison with the 
result for chains shows (fig.~\ref{fig3}) that $0<\eta<1$. As a 
consequence, the internal energy of a $N$-network can be 
approximated by $-\epsilon_0 N +\epsilon_1 c_1 N^\eta + \epsilon_3 
c_3 N$, where $\epsilon_3$ is the energy cost of a junction. We note that 
the linear term may be incorporated into the free energy 
per particle of the infinite cluster $\lambda (0,c_3)$.  The 
sublinear term must be accounted for through the energetic part 
of $F(N)$. Then, using eq.~(\ref{eq:rhonet}) and the simulation 
results of fig.~\ref{fig3} we may write, 
\begin{equation} 
\label{eq:rhonet2} \phi_n(N)=F_S(N)\exp\left[-\beta\epsilon_1 
c_1N^{\eta} + \beta N(\mu-\lambda(0,c_3))\right], 
\end{equation} 
where $\beta^{-1}\ln V F_S(N)$ is the excess entropy of a finite $N$-network 
($F_S(N)$ is expected to scale with some power of $N$ 
\cite{desCloizeaux}). 
 
Finally we turn to the analysis of the equation of state 
eq.~(\ref{eq:normal}). Using the approximations described previously ( 
$(\phi(N,2,0)+\phi(N,0,0))\propto (N^{\gamma -1}+N^{-3+\alpha})
\exp(\beta(\mu-\lambda_0))$ and $\phi_n$ given by eq.~(\ref{eq:rhonet2}))
and 
replacing the sums by integrals \cite{Sear2002}, eq.~(\ref{eq:normal}) 
is re-written as, 
\begin{equation} 
\label{eq:eqstat} 
\rho=\int_1^{N_p} dN 
(G_{rc}(N)+G_n(N))\exp\left[-\beta \Delta_n(N_p,\rho) N \right], 
\end{equation} 
where $G_{rc}\propto \left(N^\gamma+N^{-2+\alpha}\right) 
\exp(-\beta\Delta_0N)$ 
and  $G_{n}=NF_S(N) \exp(-c_1\beta\epsilon_1N^\eta)$. 
$\Delta_0\equiv\lambda_0-\lambda(0,c_3)$ is the difference 
between the free energy per particle of an infinite chain (or ring) 
and that of an infinite network. $\Delta_n\equiv\lambda(0,c_3)
-\mu(N_p,\rho)$ 
is the difference 
between the free energy per particle of an infinite network 
and that of the system. 
The dependence of these free energies on $\rho^*$ 
is estimated by analysing the simulation results for the density 
distributions of chains and networks (see fig.~\ref{fig2}). According to 
eq.~(\ref{eq:rhoNgen2}) and eq.~(\ref{eq:rhonet2}), the slopes of 
$\log\phi_n(N)$  and $\log\phi(N,2,0)$ for large $N$ 
are proportional to $-\Delta_n$ and $-\Delta_n-\Delta_0$, respectively. 
These distributions exhibit, up to $\rho^*=0.1$, the same qualitative 
behavior seen in fig.~\ref{fig2}: the slopes are always negative, 
indicating that $\Delta_n>0$ and $\Delta_n+\Delta_0>0$; the slope of 
the chain distribution is always more negative than that of networks, 
meaning that $\Delta_0>0$. 
The  dependence of 
$\log \phi_{n}(N)$ on density is depicted in the 
inset of fig.~\ref{fig2}: the slope increases with $\rho^*$  
indicating that $\Delta_n$ decreases with increasing 
density. 
Therefore, one expects $\Delta_n$ to vanish at a sufficienly large 
density, 
while $\Delta_0$  remains positive. 
Assuming that these trens hold  for an infinite 
system, the phase behavior of the DHS can be predicted.

A phase transition occurs when  
$\rho_t=\int_1^\infty dN 
(G_{rc}(N)+G_n(N))$ converges. 
The analysis of the simulation results shows that 
$\Delta_0>0$ when $\Delta_n=0$ and 
$G_n(N)\propto \exp(-c_1\beta\epsilon_1N^\eta)$ (with $0<\eta<1$); 
then  $\rho_t$ converges and at this density (in the thermodynamic 
limit, at the given $m^*$) the system exhibits a transition to a phase 
where a finite fraction of particles belongs to a macroscopic cluster. 
The existence of the transition is related to the properties of the clusters 
responsible for the convergence of $\rho_t$. 
Since the internal energy per particle of the infinite network 
($-\epsilon_0+\epsilon_3c_3$) is larger than 
that of the infinite chain ($-\epsilon_0$), 
$\Delta_0>0$ implies that the 
entropy per particle of the infinite network is larger than that of the 
infinite chain. In other words, the difference between the entropies must 
balance the increase in energy due to the junctions. 
On the other hand, the increase in the number of ends stabilizes the 
infinite network, since $0<\eta<1$. 
Thus, the transition corresponds to the emergence of a phase 
with a large configurational entropy, that balances the 
loss of translational entropy and the increase of the internal energy. 
 
Using eq.~(\ref{eq:eqstat}) we find that, in the thermodynamic limit,  
$\lim_{\rho\to \rho_t^{-}} (\frac{\partial\mu}{\partial\rho})_{m^*}$ 
is positive,  
and  conclude that the transition 
is discontinuous at $m^*=2.75$ \cite{footno2}. 
A second order phase transition occurs when the second moment of the 
distribution $\phi(N)$ diverges 
at $\rho_t$, as is the case in percolation or in Bose-Einstein 
condensation. 
 
Recently, the thermodynamics of a self-assembling system of  
chains and networks (with no rings and no attractive 
interactions between clusters) was studied using a mean-field (MF) 
approximation \cite{safran2002}. The approximations used in that work for 
the distributions of chains and networks prevent a quantitative 
comparison with the results of this letter. Nevertheless we stress that 
the phase transition mechanism is the same in both approaches. 
 
Combining a detailed analysis of extensive MC simulations and a theory 
that includes non-linear clusters, we confirmed the existence and clarified 
the nature of the phase transition of the low density DHS fluid. 
We have shown that the transition is driven by the formation of a 
macroscopic network, with a large configurational entropy that overcomes 
the cost in internal energy and the loss of translational entropy of the phase 
with high connectivity as predicted by a previous MF calculation 
\cite{safran2002}. 
 
The full phase diagram of the DHS at low densities may be studied  
by generalizing the methods described in this letter. However, the 
entropy driven transition may be pre-empted by the stabilization of 
other condensed phases (ordered or not) that are not considered in 
this framework. This is unlikely to occur at the low densities 
of the simulations described here but may prevent a critical point 
to be observed. 
 
We acknowledge D.~Levesque and P.I.C.~Teixeira for a critical reading of 
this manuscript and for their valuable suggestions.


\begin{thebibliography}{99}
\bibitem{footno2}
In a second order (or continuous) phase transition, one 
has  $(\frac{\partial^2 f}{\partial\rho^2})_{m^*}\equiv  
(\frac{\partial\mu}{\partial\rho})_{m^*} =0$.
\end{thebibliography}
\end{document}